\documentclass[superscriptaddress,nofootinbib,twocolumn,amsmath,amssymb,aps,prl]{revtex4-1}

\usepackage{xcolor}
\usepackage{braket}
\usepackage{graphicx}
\usepackage{dcolumn}
\usepackage{bm}
\usepackage[mathlines]{lineno}
\usepackage[english]{babel}
\usepackage{enumerate}
\usepackage{comment}

\newcommand{\red}[1]{{\color{red}}}

\begin{document}

\def\be{\begin{equation}}
\def\ee{\end{equation}}
\def\bea{\begin{eqnarray}}
\def\eea{\end{eqnarray}}

\title{Experimental characterization of the energetics of quantum logic gates}

\author{V. Cimini}
\address{Dipartimento di Scienze, Universita' degli Studi Roma Tre,
Via della Vasca Navale, 84 - 00146 Rome, Italy}

\author{S. Gherardini}
\address{Department of Physics and Astronomy, University of Florence, via G. Sansone 1, I-50019 Sesto Fiorentino, Italy}
\address{LENS, and QSTAR, via N. Carrara 1, I-50019 Sesto Fiorentino, Italy}
\address{SISSA, via Bonomea 265, I-34136 Trieste, Italy}

\author{M. Barbieri}
\address{Dipartimento di Scienze, Universita' degli Studi Roma Tre, Via della Vasca Navale, 84 - 00146 Rome, Italy}
\address{Istituto Nazionale di Ottica - CNR, 50125, Florence, Italy}

\author{I. Gianani}
\address{Dipartimento di Scienze, Universita' degli Studi Roma Tre, Via della Vasca Navale, 84 - 00146 Rome, Italy}
\address{Dipartimento di Fisica, Sapienza Universit\`a di Roma, Piazzale Aldo Moro, 4 - 00185 Rome, Italy}

\author{M. Sbroscia}
\address{Dipartimento di Scienze, Universita' degli Studi Roma Tre, Via della Vasca Navale, 84 - 00146 Rome, Italy}
\address{Dipartimento di Fisica, Sapienza Universit\`a di Roma, Piazzale Aldo Moro, 4 - 00185 Rome, Italy}

\author{L. Buffoni}
\address{Department of Physics and Astronomy, University of Florence, via G. Sansone 1, I-50019 Sesto Fiorentino, Italy}
\address{Department of Information Engineering, University of Florence, via S. Marta 3, I-50139 Florence, Italy}

\author{M. Paternostro}
\address{School of Mathematics and Physics, Queen's University, Belfast BT7 1NN, United Kingdom}

\author{F. Caruso}
\address{Department of Physics and Astronomy, University of Florence, via G. Sansone 1, I-50019 Sesto Fiorentino, Italy}
\address{LENS, and QSTAR, via N. Carrara 1, I-50019 Sesto Fiorentino, Italy}

\begin{abstract}
We characterize the energetic footprint of a two-qubit quantum gate from the perspective of non-equilibrium quantum thermodynamics. We experimentally reconstruct the statistics of energy and entropy fluctuations following the implementation of a controlled-unitary gate, linking them to the performance of the gate itself and the phenomenology of Landauer principle at the single-quantum level. Our work thus addresses the energetic cost of operating quantum circuits, a problem that is crucial for the grounding of the upcoming quantum technologies\textbf{}. 
\end{abstract}
 

\maketitle

Thermodynamics was developed in the $19^{\rm th}$ Century to improve the efficiency of steam engines. Its impact fostered the Industrial Revolution and affected fundamental science, technology and everyday life alike. In the $3^{\rm rd}$ Millennium, we are facing a potentially equally revolutionary process, whereby standard information technology -- usually CMOS-based -- is complemented and enhanced by quantum technologies for communication, computation and sensing~\cite{preskill2018quantum,cappellaro2017sensing}. 

Despite the significant progress made towards the implementation of prototype quantum devices able to process an increasing amount of information in a reliable and reproducible manner\,\cite{preskill2018quantum,troyer2014speedup}, little work has been devoted to the characterization of the energetic footprint of such potentially disruptive quantum technologies\,\cite{deffner2019qtd}. Yet, this is a crucial point to address: only by ensuring that the energy consumption associated with the performance of quantum information processing\,\cite{likharev1982energy, parrondo2015thermo} scales favourably with the size of a quantum processor, would the craved quantum technologies embody a credible alternative to CMOS-based devices. 

Remarkably, the fast-paced miniaturization process enabled by research in quantum information processing opens realistic possibilities to engineer and implement miniature-scale quantum machines akin to standard (macroscopic) engines that process and transform energy\,\cite{pekola2015thermo,passos2019machine}. The challenge in this respect is to make use of the emerging field of quantum thermodynamics, which aims at establishing a framework for the thermodynamics of quantum processes and systems, to design energy-efficient quantum machines\,\cite{Scully03Science299,Kosloff14ARPC65,Uzdin15PRX5,Campisi15NJP17} possibly able to outperform their classical counterparts\,\cite{troyer2014speedup}, or benchmark the performance of quantum devices from a thermodynamic perspective, as recently done for the interesting case of quantum annealears~\cite{Gardas2018,Buffoni2020}.

In this article we perform a step towards the characterization of the energetics of quantum computation by studying the energy and entropy distributions\,\cite{Albash13PRE88,Sagawa2014,Gherardini_heat,Hernandez2019,BatalhaoPRL,Gherardini_entropy,ManzanoPRX,Irreversibility_chapter} of two-qubit quantum systems realizing quantum gates. We consider the so-called two-point measurement (TPM) approach\,\cite{TalknerPRE2007} to the reconstruction of the energy and entropy generated during and as a result of the performance of a two-qubit gate. We present the experimental inference of such quantities by means of a linear-optics setting where qubits are embodied by the polarization of two photons. We then use the information gathered through the reconstructed statistics of energy and entropy distributions to assess Landauer principle at the individual-quantum level, and thus explore the relation between information processing realized through prototype two-qubit gates and the thermodynamics of such transformations. While our quantitative analysis is specific of the chosen experimental platform, our approach is based on the estimation of joint probability distributions. As such, our formal approach to the characterisation of the energetics of quantum gates would be applicable to any physical platform for quantum computation and might embody an energetics-inspired methodology for the comparison between devices implemented in different settings.

Our work embodies one of the first attempts at systematically linking the energetics of quantum information carriers to the logic functionality of a quantum gate. When developed to address multipartite settings and high-dimensional quantum systems\,\cite{daley2014many}, our approach will be pivotal to the enhancement of the performance of quantum information processes.

\section{Results}\label{results}

\subsection{The implemented process and its non-equilibrium thermodynamic analysis}

\begin{figure}[b!]
    \centering
    \includegraphics[width = \columnwidth]{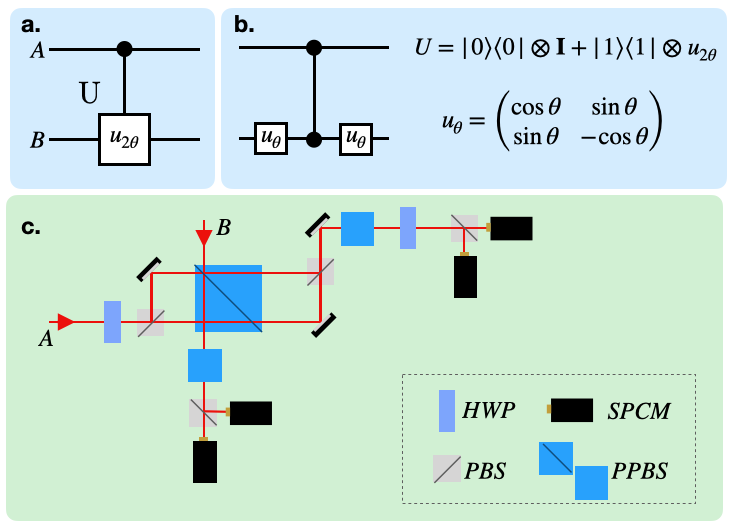}
    \caption{{\bf Details of the quantum circuit}. a. The element we inspect is a two-qubit controlled operation. b. This can be decomposed as a series of three operations: a single-qubit gate $u_\theta$, a two-qubit control-$\sigma^z$ gate, and a second gate $u_\theta$. This relies on the fact that $u_\theta u_\theta = \openone$, i.e.\,the identity, while $u_\theta  \sigma^z u_\theta = u_{2\theta}$. c. The circuit is implemented in a photonic platform with polarisation encoding. The two qubits are represented by the polarisation of photons from a standard down conversion source (not shown). The basic element of the gate is a Partially Polarising Beam Splitter (PPBS) with expected transmittivities $T_H=1$, $T_V=1/3$ for the two polarisations. This is embedded in a polarisation Mach-Zehnder to mitigate the effects of the deviation of $T_H$ from 1 (we have $T_H \simeq 0.985$ in our experiment) - for the sake of phase stability this was implemented as a Sagnac loop. Further PPBSs are needed in order to equalise losses on the two polarisations. Half-wave plates (HWP) set at an angle $\theta/2$ are used to implement $u_\theta$. Polarisation measurements in the logic basis are implemented using polarising beam splitters (PBSs) and single-photon counting modules (SPCMs).}
    \label{fig:scheme}
\end{figure}

We consider the simple quantum circuit in Fig.\,\ref{fig:scheme}: our two-qubit gate performs a controlled-unitary $\textsc{U}$, based on its decomposition into local unitary gates $u_\theta$ and a control-$\sigma^z$ gate, that applies a Pauli $\sigma^{z}$ gate to the state of qubit $B$, conditioned on the state of the qubit $A$. This is the simplest instance of a programmable quantum circuit with a two-qubit interaction, an essential feature to our purposes.  

Our physical implementation, depicted in Fig.\,\ref{fig:scheme}c, adopts  the two-photon control-$\sigma^z$ gate based on the use of polarisation-selective non-classical interference and post-selection~\cite{Langford05, Kiesel05, Okamoto05}. The polarisation encoding represents the logical states $\vert 0\rangle_k$ and $\vert 1\rangle_k$ with the horizontal $\vert H\rangle_k$ and vertical $\vert V\rangle_k$ polarisations, respectively, for both qubits $k=A$ and $B$.

In order to investigate the thermodynamics associated with the performance of our gate, we need to associate an Hamiltonian to the interacting system of the two qubits able to account for the action of the device. To this purpose, we introduce the total Hamiltonian $\mathcal{H}_{\rm tot}=\mathcal{H}_{L}+\mathcal{H}_{\rm int}$, which consists  of the local Hamiltonian $\mathcal{H}_{L}$ and the interaction term $\mathcal{H}_{\rm int}$, which read
\begin{equation}\label{eq:eq1}
\begin{aligned}
\mathcal{H}_L &= {\frac{1}{2}}\hbar\,\omega_{L}\left( \sigma_A^z \otimes \openone_B + \openone_A \otimes \sigma_B^z\right),\\
\mathcal{H}_{\rm int} &= \frac12\hbar\,\omega_{\rm int}|1\rangle\!\langle 1|_A\otimes\sigma_B^x \,.
\end{aligned}
\end{equation}
In Eq.\,(\ref{eq:eq1}), $\openone$ is the identity matrix and $\sigma_k^j$ is the $j=x,y,z$ Pauli matrix of the qubit $k$. The interaction Hamiltonian ${\cal H}_{\rm int}$ generates a rotation of the state of qubit $B$ that is conditioned on the state of qubit $A$. In particular, the total Hamiltonian ${\cal H}_{\rm tot}$ generates the following trajectories for the two-qubit logical states
\begin{equation}
\label{eq:HH--VV-VH_VV}
\begin{aligned}
  |0\rangle_A  |0\rangle_B  &\rightarrow e^{i\omega_L t}|0\rangle_A  |0\rangle_B,\\
  |0\rangle_A  |1\rangle_B  &\rightarrow |0\rangle_A  |1\rangle_B, \\
  |1\rangle_A  |0\rangle_B  &\rightarrow e^{-i \omega_L t/2}\left(h_{1}(t)|1\rangle_A  |0\rangle_B + h_{2}(t)|1\rangle_A  |1\rangle_B \right), \\
  |1\rangle_A  |1\rangle_B  &\rightarrow e^{-i \omega_L t/2}\left(h_{2}(t)|1\rangle_A |0\rangle_B + h^*_{1}(t)|1\rangle_A  |1\rangle_B\right),
\end{aligned}
\end{equation}
with
\begin{equation}
h_1(t)= \cos\left(\Delta t\right)+i\frac{\omega_{L}}{2\Delta}\sin\left(\Delta t\right),\quad h_2(t)= -i\frac{\omega_{\rm int}}{2\Delta}\sin\left(\Delta t\right)
\end{equation}
and $\Delta=\sqrt{\omega_{L}^2 + \omega_{\rm int}^2}/2$. The unitary operation ${\cal U}_{AB}(t)$ accounting for the trajectories in Eq.\,\eqref{eq:HH--VV-VH_VV} can be cast in the following form
\begin{equation}
\label{U}
{\cal U}_{AB}(t)=e^{i\omega_Lt}|0\rangle\!\langle0|_A\otimes{\cal P}_B(t)+e^{-i\omega_Lt/2}|1\rangle\!\langle1|_A\otimes{\cal R}_B(t)
\end{equation}
with ${\cal P}_B(t)\equiv{\rm diag}[1,e^{-i\omega_{L}t}]$ a phase gate on qubit $B$ and ${\cal R}_B(t)$ a single-qubit rotation of a time-dependent angle $\varphi=\Delta t$ around an axis identified by the vector ${\bf n}=(\sin\zeta,0,\cos\zeta)$,
where $\zeta=\cos^{-1}(\omega_L/2\Delta)$. The transformations in Eq.\,\eqref{eq:HH--VV-VH_VV} thus correspond to those imparted by our gate, up to phases  $\cos\theta = \vert h_1(t) \vert$, $\sin \theta = \vert h_2(t)\vert$ that, as we will see, do not influence the energetics of the process. We can then rely on our model to analyse the computation processes from an out-of-equilibrium thermodynamic perspective.

\subsection{Energy and entropy distributions}

The framework for the inference of the statistics of energetics arising from Eq.\,\eqref{U} relies on two crucial points 
\begin{itemize}
\item[{\bf 1.}] We adopt the tool provided by the TPM scheme to characterize energy and entropy distributions. This implies the application of two projective measurements onto the {energy eigenstates} of the two-qubit system at the initial and final  instants of time of the evolution of the system. 
\item[{\bf 2.}] We assume to apply only \emph{local} energy measurements. We will thus be unable to access quantum correlations between $A$ and $B$.
\end{itemize}

In light of the first energy measurement entailed by the TPM approach, any quantum coherence potentially present in the initial state of the system is destroyed. While this is an intrinsic feature of the chosen approach to the inference of energy fluctuations,  recently alternative methodologies have been developed which enable the retention of the effects of initial quantum coherences~\cite{Allahverdyan2014,Miller2017,Levy2019,Landi2019,Gherardini2020,Diaz2020}. The use of such approaches in the context of our investigation will be reported elsewhere.

Point {\bf 2} has a deep implication: One has access only to the energy values pertaining to the local Hamiltonian of the two qubits. This means that, at the end of the protocol, the energy of $A$ and $B$ will be one of the eigenvalues of the Hamiltonian terms $\mathcal{H}_{L_{A}} \equiv \hbar \omega_{L}(\sigma_A^z \otimes \mathbb{I})/2$ and $\mathcal{H}_{L_{B}} \equiv \hbar \omega_{L}(\mathbb{I} \otimes \sigma_B^z)/2$, i.e., $E_{j_{k}}$~($j=0,1$ and $k=A,B$). In order to realise this, we introduce the local measurement operators 
\begin{equation}
\Pi_{\psi_{A}\phi_{B}} \equiv |\psi\rangle\!\langle\psi|_{A} \otimes |\phi\rangle\!\langle\phi|_{B}
\end{equation}
with $\psi,\phi=0,1$. Each of the four projectors $\Pi_{\psi_{A}\phi_{B}}$ is associated with the corresponding energy value $E_{\psi_{A}\phi_{B}} = E_{\psi_A} + E_{\phi_B}$. Those values correspond to stochastic realizations of the composite system at a given time-instant $t$. We should stress how these stochastic realizations $E_{\psi_{A}\phi_{B}}$ do not contain any correlation originated by the interaction Hamiltonian $\mathcal{H}_{\rm int}$. In what follows we will express the values of $E_{\psi_{A}\phi_{B}}$ in units of $\hbar\omega_{L}$, so that $E_{\psi_{A}\phi_{B}}\in\{-2,0,2\}$.

The energy variations are then observed under the lens of the probability distributions associated to the stochastic energy changes [cf.\,Appendices~A and B for the formal definition of energy and entropy production]. In the chosen computational basis and by encoding the binary value of $\psi_{A}\phi_{B}$ in the integer $m$ so that $E_{\psi_{A}\phi_{B}}\equiv E_m$, the system energy variation reads $\Delta E_{m,n} \equiv E_{m}^{\rm fin} - E_{n}^{\rm in}$\,\cite{TalknerPRE2007,Gherardini_heat}. Thus, $E_{n}^{\rm in}$ ($E_{m}^{\rm fin}$) is the measured energy of the quantum system at the initial (final) time $t_0$ ($t_{\rm fin}$).

The probability distribution of the energy variations $\Delta E$ can be then formally written as
\begin{equation}
{\rm Prob}(\Delta E)=\sum_{m,n}\delta(\Delta E - \Delta E_{m,n})p(E_{n}^{\rm in},E_{m}^{\rm fin})\,,
\end{equation}
where $\delta(x)$ is the Kronecker delta with argument $x$, and $p(E_{n}^{\rm in}) = p(E_{\psi_A\phi_B}^{\rm in}) \equiv {\rm Tr}[\rho_{0}\,\Pi_{\psi_A\phi_B}^{\rm in}]$ is the probability that the initial value of energy is $E^{\rm in}_n$. Moreover,
\begin{equation}\label{joint_p}
p(E_{n}^{\rm in},E_{m}^{\rm fin}) = {\rm Tr}\left[\Pi_{\psi_A\phi_B}^{\rm fin}U_{t_{\rm fin}}\Pi_{\psi_A\phi_B}^{\rm in}U_{t_{\rm fin}}^{\dagger}\right]p(E_{n}^{\rm in})
\end{equation}
denotes the joint probability to measure $E_{n}^{\rm in}$ at $t_0$ and $E_{m}^{\rm fin}$ at $t_{\rm fin}$ by performing local measurements at time $t$, whereby $\Pi^t_{\psi_A\phi_B} \equiv \Pi^t_{\psi_A}\otimes\Pi^t_{\phi_B}$. More details can be found in the Methods section. Notice that we have denoted as $\rho_0$ the initial state of the system before the first measurement of the TPM scheme is performed. The initial state $\rho_0$ can be an arbitrary density operator, including the case of initial states with quantum coherence in the energy (Hamiltonian) basis of the system. This means that, by applying a TPM scheme, all the information about $\rho_0$ is contained in the probability $p(E_{k}^{\rm in})$\,\cite{Perarnau-LlobetPRL2017}. Finally, it is worth observing that experimentally (as it will shown below) the conditional probabilities $p(E_{n}^{\rm in},E_{m}^{\rm fin})/p(E_{n}^{\rm in}) \equiv p(E_{n}^{\rm in}|E_{m}^{\rm fin})$ can be obtained by initializing the system in the product states $\Pi_{\psi_A\phi_B}^{\rm in}$, letting it evolve according to ${\cal H}_{\rm tot}$ and then measuring its energy at final time $t_{\rm fin}$.

\subsection{Experimental characterisation}

In the previous section we have established a connection between the action of our gate in Fig.\,\ref{fig:scheme} and the dynamics encompassed in Eq.\,\eqref{eq:HH--VV-VH_VV}; we can thus proceed to its characterization.

We first address the energy fluctuations observed in the local energy basis. The specific structure of Eq.\,\eqref{eq:HH--VV-VH_VV} imposes stringent constraints to the values assumed by the conditional probabilities $p(E_{m}^{\rm fin}|E_{n}^{\rm in})$. In particular, one finds that the only conditional probabilities different from zero are 
$p(E_{0_{A}0_{B}}^{\rm fin}|E_{0_{A}0_{B}}^{\rm in})=p(E_{0_{A}1_{B}}^{\rm fin}|E_{0_{A}1_{B}}^{\rm in})=1$ and 
\begin{equation}\label{eq:cond_prob_exp}
\begin{aligned}
&p(E_{1_{A}0_{B}}^{\rm fin}|E_{1_{A}0_{B}}^{\rm in}),\quad p(E_{1_{A}0_{B}}^{\rm fin}|E_{1_{A}1_{B}}^{\rm in}),\\
&p(E_{1_{A}1_{B}}^{\rm fin}|E_{1_{A}0_{B}}^{\rm in}),
\quad p(E_{1_{A}1_{B}}^{\rm fin}|E_{1_{A}1_{B}}^{\rm in}),
\end{aligned}
\end{equation}
which are all functions of $h_1(t)$ and $h_2(t)$. Notice that the trajectories associated with the conditional probabilities in Eq.\,\eqref{eq:cond_prob_exp} leave qubit $A$ in the logical state $|1\rangle_A$ and modify the state of qubit $B$. As first step, we have thus experimentally reconstructed the conditional probabilities using the experimental apparatus in Fig.\,\ref{fig:scheme}.
\begin{figure}[b!]
    \centering
    \includegraphics[width = \columnwidth]{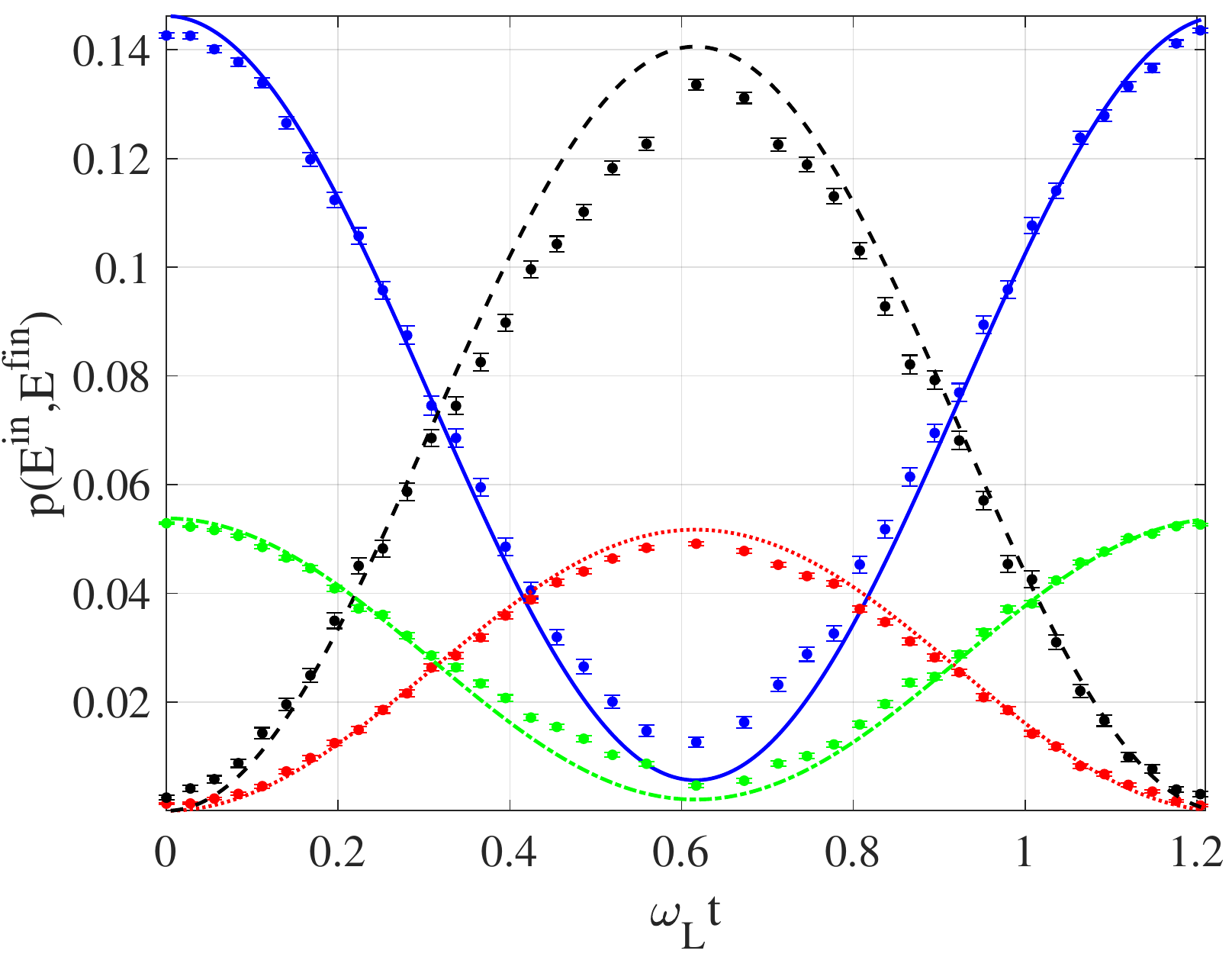}
    \caption{{\bf Joint probabilities}. Comparison between theoretical (lines) and experimental (dots) joint probabilities $p(E_{1_{A}\phi_{B}}^{\rm in},E_{1_{A}\phi_{B}}^{\rm fin})$ as a function of time $t$, with $\omega_{\rm int}/\omega_L = 5$ in natural units, $\rho_0$ as in Eq.~(\ref{eq_rho_0}) and $|V\rangle_k=(1,0)$, $|H\rangle_k=(0,1)$ for both qubits $k=A$ and $B$. We show $p(E^{\rm in}_{1_{A}0_{B}},E^{\rm fin}_{1_{A}0_{B}})$ (blue dots and solid line), $p(E^{\rm in}_{1_{A}1_{B}},E^{\rm fin}_{1_{A}0_{B}})$ (red dots and dotted line), $p(E^{\rm in}_{1_{A}0_{B}},E^{\rm fin}_{1_{A}1_{B}})$ (black dots and dashed line), and $p(E^{\rm in}_{1_{A}1_{B}},E^{\rm fin}_{1_{A}1_{B}})$ (green dots and dot-dashed line).}
    \label{fig:joint_prob}
\end{figure}
In this regard, in Fig.\,\ref{fig:joint_prob} we plot the comparison between the joint probabilities $p(E_{1_{A}\phi_{B}}^{\rm in},E_{1_{A}\phi_{B}}^{\rm fin})$ as obtained by numerical simulations and the analysis of the experimental data. As $p(E_{k}^{\rm in})$ depends on the specific choice of the initial state $\rho_0$ of the two-qubit system, also the joint probabilities in Fig.\,\ref{fig:joint_prob} would bear a dependence on the value taken for $\rho_0$. Although, in principle, any choice of $\rho_0$ would be equally valid, the test of Landauer principle reported later requires a \emph{thermal} initial density operator. We have thus considered
\begin{equation}\label{eq_rho_0}
\rho_0 = \bigotimes_{k=A,B} \frac{e^{-\beta_k\omega_{L}\sigma_k^z}}{{\rm Tr}[e^{-\beta_k\omega_{L}\sigma_k^z}]}
\end{equation}
where $\beta_A=\frac{1}{2\omega_L}\ln\frac{\alpha}{1-\alpha}$, with $\alpha\in[0,1]$, and $\beta_B = 1/(2\omega_L)$. In our case, we have chosen $\alpha=0.2$, so as to ensure a prominent asymmetry among the populations of qubit $A$.  
\begin{figure}[t!]
    \centering
    \includegraphics[width = \columnwidth]{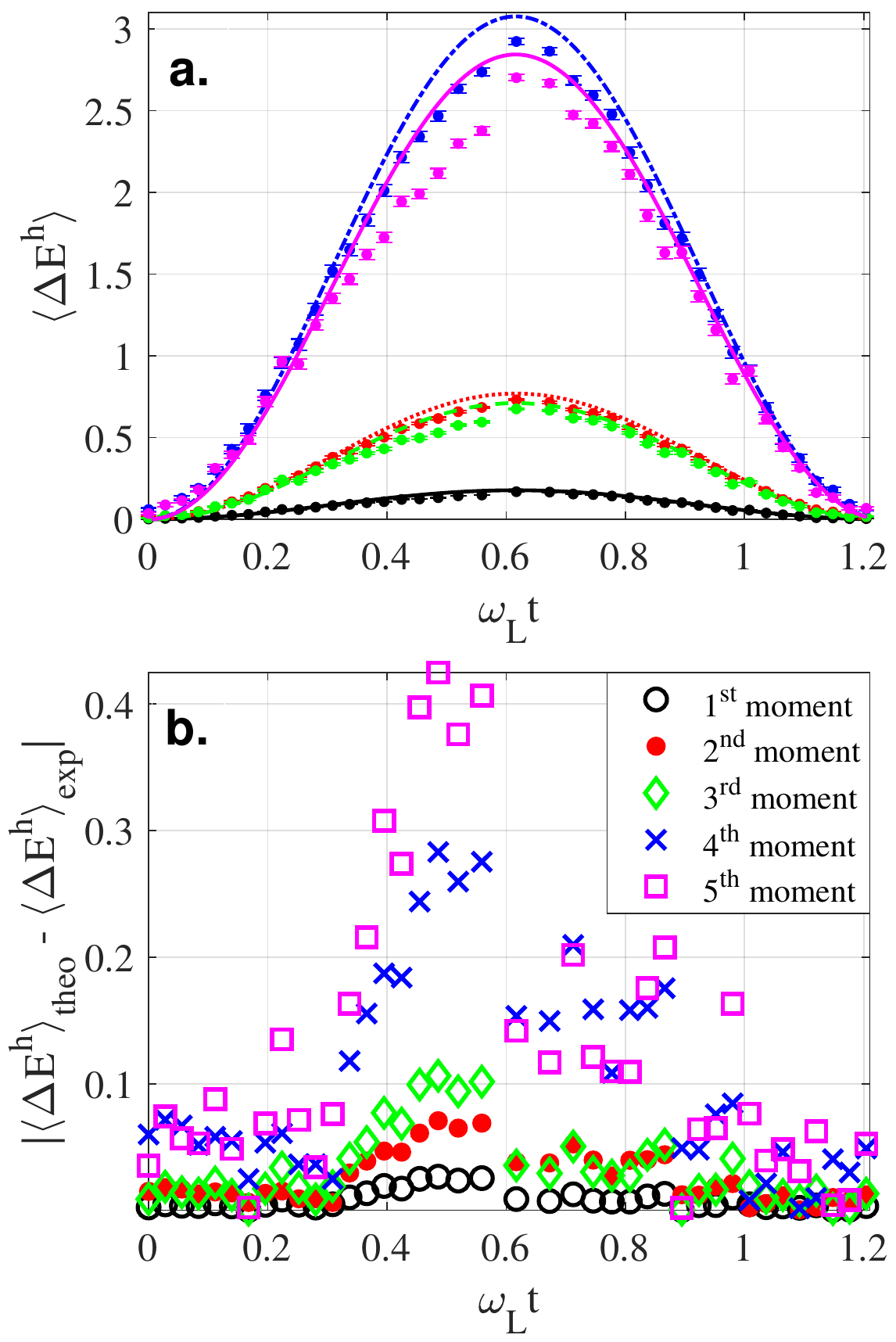}
    \caption{{\bf Statistical moments of $\Delta E$}. a. Statistical moments of the energy probability distribution as a function of time. We compare the experimental values and the corresponding theoretical predictions. The initial state of the two-qubit system is the density matrix $\rho_0$ in Eq.\,(\ref{eq_rho_0}). We show the experimental (theoretical) values of the  moments from $1^{\rm st}$ to $5^{\rm th}$ as black, red, green, blue, magenta dots (lines) for the same parameters as in Fig.\,\ref{fig:joint_prob}.
    b. Absolute error $|\langle\Delta E^h\rangle_{\rm theo}-\langle\Delta E^h\rangle_{\rm exp}|$ accounting for the difference between theoretical and experimental statistical moments of $\Delta E$ from $1^{\rm st}$ to $5^{\rm th}$, provided by black circles, red dots, green diamonds, blue crosses and magenta squares, respectively.}
    \label{fig:heat_stat_moments}
\end{figure}
Eq.~\eqref{eq_rho_0} has been simulated measuring single-photon orthogonal polarisation states for a time-interval proportional to the probability that such polarization state occurs in $\rho_0$. The desired state can be obtained by mixing the weights of the two polarization states.

In Fig.\,\ref{fig:heat_stat_moments}a we plot the first $5$ statistical moments $\langle\Delta E^{h}\rangle$, $h=1,..,5$, of the probability distribution pertaining to the energy variation during the implemented process, i.e., 
\begin{equation}
\langle\Delta E^{h}\rangle \equiv \sum_{n,m}p(E_{n}^{\rm in},E_{m}^{\rm fin})\Delta E_{m,n}^{h} \ ,
\end{equation}
depending on the evolution time $t$. From Fig.\,\ref{fig:heat_stat_moments}, one can observe that all the energy statistical moments have a periodic time-behaviour with maximum values in correspondence of $\omega_{L}t = k\pi/\sqrt{26}\approx 0.62$, with $k$ integer number. Moreover, being the error in the experimental data larger at $\omega_{L}t = k\pi/\sqrt{26}$, it is worth also noting that the curves of $\langle\Delta E^{h}\rangle$ obtained by the experiment with $h \geq 4$ lose in accuracy in such points. This generally holds also for the other figures. Deviations of the experimental data from theoretical expectations are mostly due to two factors. The first is the non-ideal visibility of quantum interference due to the differences between the effective values of the transmittivity of the PPBS used to implement our gate from the ideal one (cf. caption of Fig.\,\ref{fig:scheme}). The second is the presence of random accidental counts, which decrease this visibility even further. In terms of energetics, these imperfections are reflected in a smaller values of the observed high-order energy moments close to the maximum achieved at $\omega_L t\simeq0.62$, which turns out to be a very sensitive working point. The energy lost to the environment limits the extent of the fluctuations. 

Any non-equilibrium process results in the production of irreversible entropy $\Delta\sigma$\,\cite{Irreversibility_chapter}, which embodies a thermodynamic quantifier of the break-down of time-reversal symmetry\,\cite{BatalhaoPRL} and a non-equilibrium restatement of the second law of thermodynamics. At the quantum level, its values are determined by the competition of two sources of randomness: a classical one associated with the choice of initial state [cf.\,Eq.\,\eqref{eq_rho_0}] and a quantum mechanical one induced by the stochastic nature of quantum trajectories, which renders entropy production an inherently aleatory quantity. 

One can thus introduce the associated probability distribution for the quantum entropy production\,\cite{Irreversibility_chapter,BatalhaoPRL} to study the statistics of such quantity. As illustrated in methods, for any unitary dynamical evolution, $\Delta\sigma_{m,n} \equiv \Delta\sigma(E^{\textrm{fin}}_{m},E^{\textrm{in}}_{n})$ obeys a quantum fluctuation theorem\,\cite{Gherardini_entropy,ManzanoPRX,Irreversibility_chapter} so that
\begin{equation}
\Delta\sigma_{m,n} = \ln [p(E^{\textrm{in}}_{n})/p(E^{\textrm{fin}}_{m})] \ ,
\end{equation}
where $p(E^{\textrm{in}}_{n})$ and $p(E^{\textrm{fin}}_{m})$ denote the probabilities to measure the $n^{\rm th}$ and $m^{\rm th}$ energy outcome at $t_0$ and $t_{\rm fin}$, respectively. In particular, $p(E^{\textrm{fin}}_{m})$ is given by the following relation:
\begin{eqnarray}
p(E^{\textrm{fin}}_{m}) &\equiv& {\rm Tr}[\rho_{\rm fin}\,\Pi_{\psi_A\phi_B}^{\rm fin}]\nonumber \\
&=& {\rm Tr}[\mathcal{U}_{AB}(t_{\rm fin})\,\rho_{\rm in}\,\mathcal{U}^{\dagger}_{AB}(t_{\rm fin})\Pi_{\psi_A\phi_B}^{\rm fin}] \ ,
\end{eqnarray}
where $\rho_{\rm fin}$ is the density operator of the bipartite quantum system at the end of its dynamical evolution. $\rho_{\rm fin}$ can be experimentally obtained by preparing the quantum system in the ensemble average
\begin{equation}
\rho_{\rm in} = \sum_{\psi_A\phi_B}p(E^{\textrm{in}}_{\psi_A\phi_B})\Pi_{\psi_A\phi_B}^{\rm in}
\end{equation}
after the $1^{\rm st}$ energy measurement of the TPM scheme and then letting it evolve. Since the unitary operator $\mathcal{U}_{AB}(t)$ acts separately on qubits $A$ and $B$ by means of product state operations, also the final probability $p(E^{\textrm{fin}}_{m})$ as well as the $16$ realizations of the stochastic quantum entropy production have been obtained by experimental data. Further details are in methods section.   

\begin{figure}[t!]
    \centering
    \includegraphics[scale = 0.63]{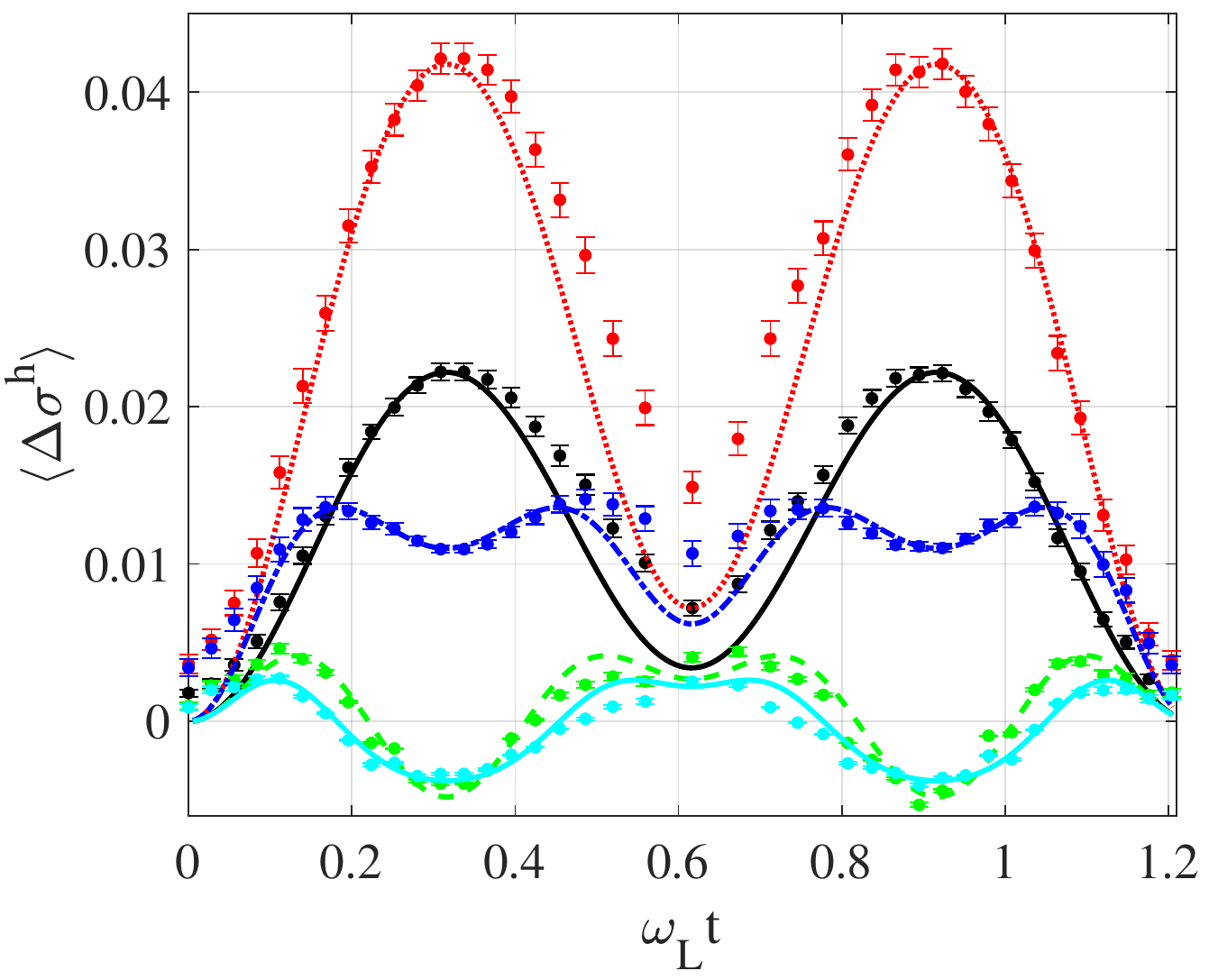}
    \caption{{\bf Statistical moments of $\Delta\sigma$}. Statistical moments of the entropy probability distribution as a function of time: Comparison between experiment (dots) and theory (lines). As in the other figures,  $\omega_{\rm int}/\omega_L = 5$ with $\rho_0$ given by Eq.~(\ref{eq_rho_0}). We show the experimental (theoretical) values of the moments from $1^{\rm st}$ to $5^{\rm th}$ as  black, red, green, blue, and cyan dots (lines).}
    \label{fig:entropy_moments}
\end{figure}

\begin{figure}[t!]
    \centering
    \includegraphics[width = 0.95\columnwidth]{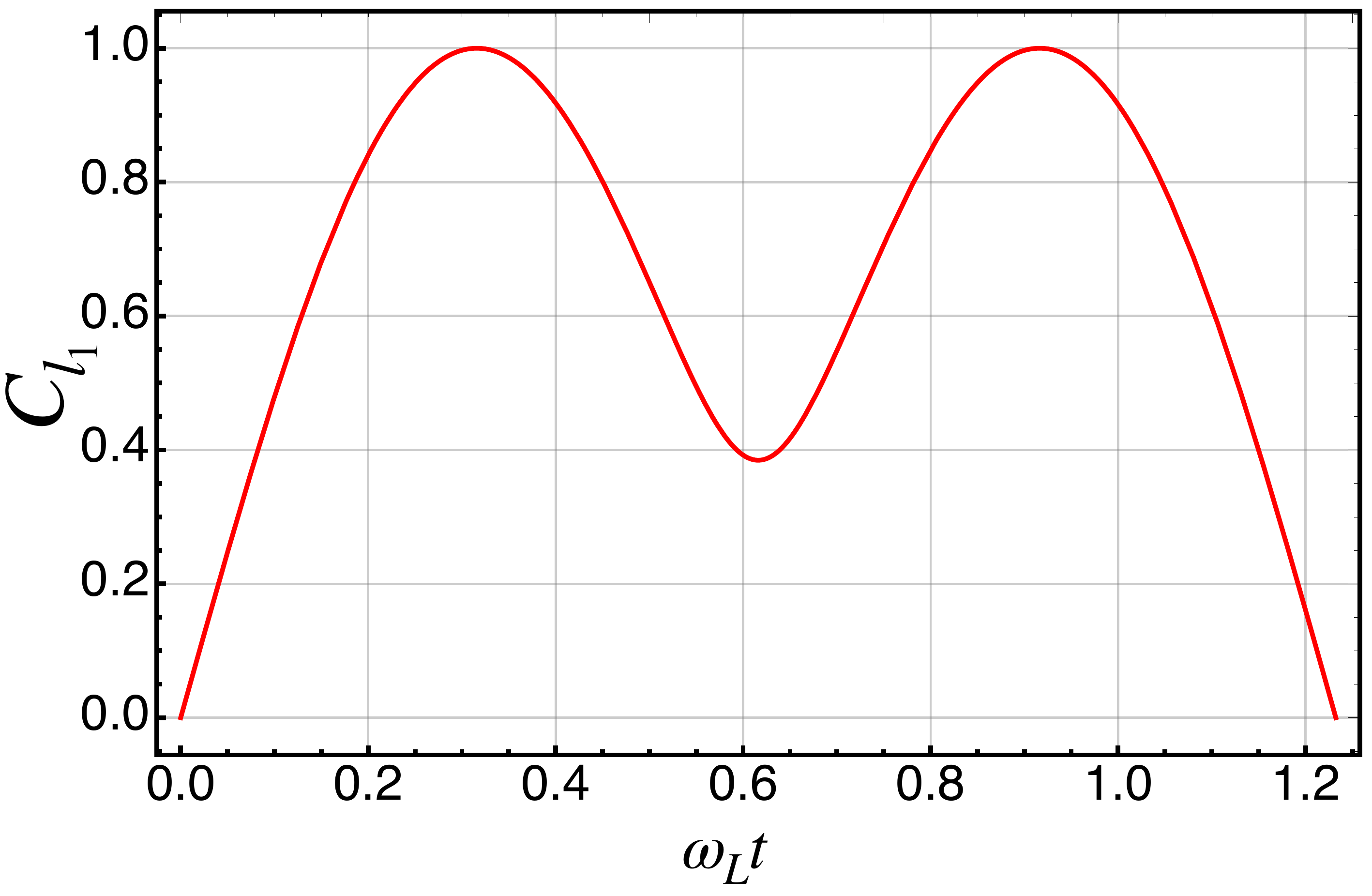}
    \caption{{\bf Quantum coherence measure}. Plot of the $l_1$-norm of quantum coherence against the dimensionless evolution time $\omega_L t$ for the same parameter as in the other figures and assuming the initial state $|0\rangle_A|1\rangle_{B}$ or $|1\rangle_A|0\rangle_{B}$.}
    \label{fig:coherence}
\end{figure}

The statistics of $\Delta\sigma$ is determined by evaluating the corresponding probability distribution $\textrm{Prob}(\Delta\sigma)$. As shown in methods, this is given by 
\begin{equation}\label{prob_sigma}
\textrm{Prob}(\Delta\sigma) = \sum_{n,m}\delta\left(\Delta\sigma - \Delta\sigma(E^{\textrm{in}}_{n},E^{\textrm{fin}}_{m})\right)p(E_{n}^{\rm in},E_{m}^{\rm fin}) \ ,
\end{equation}
where the joint probabilities $p(E_{n}^{\rm in},E_{m}^{\rm fin})$ are the same of those used to derive the energy probability distribution. Thus, having experimentally measured the joint probabilities $p(E_{n}^{\rm in},E_{m}^{\rm fin})$ [cf.\,Fig.\,\ref{fig:joint_prob}] and then obtained the stochastic realizations of $\Delta\sigma$, we can directly derive the statistical moments of the entropy distribution. In Fig.\,\ref{fig:entropy_moments} we plot the comparison between the experimental and theoretical statistical moments 
\begin{equation}
\langle\Delta\sigma^{h}\rangle \equiv \sum_{n,m}p(E_{n}^{\rm in},E_{m}^{\rm fin})\Delta\sigma_{m,n}^{h} \ .
\end{equation}
The experimental evidence shows a good agreement with the theoretical predictions. Moreover, an important  physical point can be drawn: the black line and dots in Fig.\,\ref{fig:entropy_moments}, which show the behavior of the average stochastic entropy production, closely resembles the trend followed by the {\it $l_1$-norm of quantum coherence}\,\cite{Baumgratz}
\begin{equation}
    C_{l_1}(\rho)=\sum_{i\neq j}|\rho_{ij}|
\end{equation}
with $\rho_{ij}$ denoting the $(i,j)$ entry of the density matrix $\rho$. When evaluated for the last two trajectories in Eq.\,\eqref{eq:HH--VV-VH_VV}, $C_{l_1}$ follows the same trend as shown in Fig.\,\ref{fig:coherence}. The stationary points  of $C_{l_1}$ achieved at $\omega_L t=k\pi/\sqrt{26}~(k=1,2,3)$ correspond to analogous extremal points for $\langle\Delta\sigma\rangle$, thus corroborating the expectation that dynamically created quantum coherence play a crucial role in the determination of the amount of irreversible entropy generated across a non-equilibrium process\,\cite{Santos2018,Francica2019}. As a matter of fact, quantum coherence embodies an additional source of entropy production that adds to the (classical) contribution provided by the populations of the density matrix under analysis. The experimental simulation of a non-equilibrium process discussed in this paper provides a striking instance of such interplay between classical and genuinely quantum contributions to entropy production and strikingly connects them to the functionality of the gate encompassing the dynamics that we have addressed.

Such connections are reinforced by the assessment of a Landauer-like relation connecting stochastic energetics between qubit $A$ and $B$ and the associated entropy production, which can be cast as\,\cite{ReebNJP2014,GooldPRL2015,YanPRL2018} 
\begin{equation}
\beta\langle\Delta E\rangle \geq \langle\Delta\sigma\rangle\,,
\end{equation}
where $\beta$ is the inverse temperature of the initial reduced state of qubit $B$.
\begin{figure}[b!]
    \centering
    \includegraphics[width = \columnwidth]{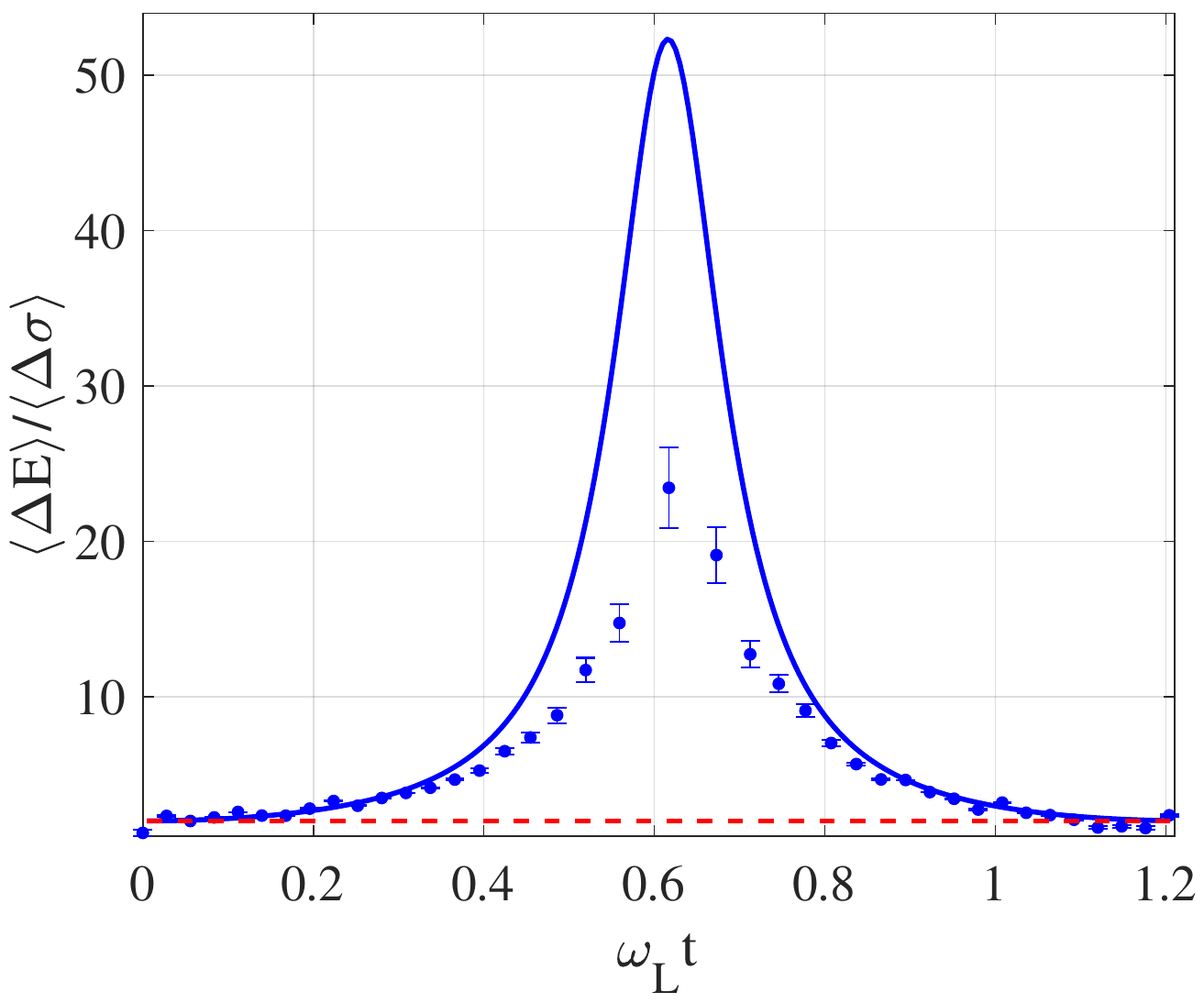}
    \caption{{\bf Landauer principle}. Ratio between $\langle\Delta E\rangle$ and $\langle\Delta\sigma\rangle$: Comparison between experiment (dotted blue line) and theory (solid blue line), with $\omega_{\rm int}/\omega_{L} = 5$ (in natural units) and $\rho_0$ thermal state (with $\beta = 1/2$) of Eq.\,(\ref{eq_rho_0}). The Landauer principle $\beta\langle\Delta E\rangle \geq \langle\Delta\sigma\rangle$ ($1/\beta$, red dashed line) is respected both theoretically and experimentally, although the experimental dataset underestimates the energy-to-entropy ratio close to the maximum achieved at $\omega_L t\simeq 0.62$ since experimental value of $\langle\Delta\sigma\rangle$ at such time deviates from the nearly null expected one}.
    \label{fig:Landauer}
\end{figure}
Fig.\,\ref{fig:Landauer} compares the ratio  $\langle\Delta E\rangle/\langle\Delta\sigma\rangle$ with $1/\beta$, showing full agreement with the predictions of Landauer principle and demonstrating an enhanced energy-to-entropy trade-off at the time minimizing the entropy production. The experimental dataset underestimates the energy-to-entropy ratio close to the maximum achieved at $\omega_L t\simeq0.62$. This is due to the fact that the experimental value of $\langle\Delta\sigma\rangle$ at such time deviates from the nearly null expected one, thus lowering the observed ratio $\langle\Delta E\rangle/\langle\Delta\sigma\rangle$.

While the results reported here refer to the initial thermal state in Eq.\,\eqref{eq_rho_0}, as mentioned earlier, our general methodology for the inference of $\langle\Delta E\rangle$ and $\langle\Delta\sigma\rangle$ do not depend on the specific choice of $\rho_0$. Thus, it could be also used to measure Clausius-like relations for an arbitrary quantum system, without initializing the system in a thermal state, which would embody an interesting development of the endeavours reported here.

\section{Discussions}\label{conc}

We have performed the theoretical and experimental characterisation of the energetics of a quantum gate implementing a controlled two-qubit quantum gate. Using tools specifically designed for the quantification of energy changes resulting from a non-equilibrium  process, we have inferred the statistics of quantum energy fluctuations and entropy production resulting from the implementation of such gate in a linear-optics platform where information carriers are encoded in the polarization of two photons.\\
The inferred statistics brings about clear signatures of the influence of quantum coherence generated in the state of the two qubits by the quantum gate, and allows for the test of the energy-to-entropy trade-off embodied by Landauer principle, thus connecting in a quantitative manner logical and thermodynamic irreversibility entailed by the experimental process that we have realised. Our investigation is aligned with current efforts aimed to bridge quantum thermodynamics and quantum information processing, which hold the promises  to deliver a deeper understanding of the origin of quantum advantage, based on non-equilibrium thermodynamics.\\
Our experiment was purposely chosen as the simplest instance for the sake of clarity. It will be interesting to extend our study to recent coherence-preserving approaches to the quantification of the distribution of energy fluctuations~\cite{Allahverdyan2014,Miller2017,Levy2019,Landi2019,Gherardini2020} to check the influences that such key quantity has on the energetic footprint of the gate. Moreover, consideration should also be given to multi-qubit gates to check if the full detail of the circuital scheme needs being taken into account for the drawing of suitable thermodynamic bounds.  

\section{Methods}

\subsection{Energy change distribution}
\label{appxA}

At the nanoscale, non-equilibrium fluctuations play a crucial role and we expect that they are responsible for the \emph{irreversible} exchange of energy between the system and the external environment. In this regard, the energy change $\Delta E_{m,k} \equiv E_{m}^{\rm fin} - E_{k}^{\rm in}$ during the evolution of the system is defined as the difference between the measured energy of the quantum system, respectively before ($1^{\rm st}$ energy measurement at $t_0$) and after ($2^{\rm nd}$ energy measurement at $t_{\rm fin}$) its dynamics. The Hamiltonian of the system is assumed to be time-independent, so that the energy values that the system can take remain the same during all its dynamics: no coherent modulation of the Hamiltonian is indeed considered. For this reason, no ``mechanical'' work is produced by or on the system, with the result that all the energy variations have to be ascribed to loss into the environment in the form of heat. In case the Hamiltonian of the system is time-independent and energy changes are induced by measurement processes, one could refer to \emph{quantum-heat}~\cite{Elouard2016,Gherardini_heat}. 

As usual, the probability distribution of $\Delta E$ is given by the combinatorial combination of the Kronecker delta $\delta(\Delta E - \Delta E_{m,k})$ weighted by the joint probability $p(E_{k}^{\rm in},E_{m}^{\rm fin})$ to measure $E_{k}^{\rm in}$ at $t_0$ and $E_{m}^{\rm fin}$ at $t_{\rm fin}$. Formally, one has
\begin{equation}
{\rm Prob}(\Delta E)=\sum_{k,m}\delta(\Delta E - \Delta E_{m,k})p(E_{k}^{\rm in},E_{m}^{\rm fin}) \ .
\end{equation}
The expression of the joint probabilities $p(E_{k}^{\rm in},E_{m}^{\rm fin})$, as well as the specific values of $E_{k}^{\rm in}$ and $E_{m}^{\rm fin}$, changes whether we apply global or local energy measurements on the bipartite quantum system. Here, we will analyze only the case of applying local energy measurements, for which we just take into account the local Hamiltonian $\mathcal{H}_{L_A}$ and $\mathcal{H}_{L_B}$ of ${\rm A}$ and ${\rm B}$. They can be generally decomposed as
\begin{equation}
\mathcal{H}_{L_A} = \sum_{\psi_{A}\in\{H,V\}}E_{\psi_{A}}\Pi_{\psi_{A}}\,\,\,;\,\,\,
\mathcal{H}_{L_B} = \sum_{\phi_{B}\in\{H,V\}}E_{\phi_{B}}\Pi_{\phi_{B}}\,
\end{equation}
with $\{|H\rangle,|V\rangle\}$ computational basis of the single partition (qubit). As a remark, observe that the complete (or maximally informative) characterization of energy and entropy production of a multipartite quantum system is achieved by performing global measurements, since also the effects of (quantum) correlations between each partition are properly taken into account. However, local measurements are simpler to be performed and sometimes they constitute the only experimentally possible solution.

Hence, if only local energy measurements are performed, the joint probabilities of the distribution ${\rm Prob}(\Delta E)$ are generally provided by the following relation:
\begin{equation}
    \begin{aligned}
&p(E_{k}^{\rm in},E_{m}^{\rm fin}) \equiv
{\rm Tr}\left[(\Pi_{\psi_A}^{\rm fin}{\otimes}\Pi_{\phi_B}^{\rm fin})\Lambda_{t_{\rm fin}}[\Pi_{\psi_A}^{\rm in}{\otimes}\Pi_{\phi_B}^{\rm in}]\right]p(E_{k}^{\rm in}),
\end{aligned}
\end{equation}
where $\Lambda_{t}$ is a complete positive trace preserving (CPTP) map \cite{Caruso2014} modeling the evolution of the system, while
\begin{equation}
p(E_{k}^{\rm in}) \equiv {\rm Tr}[(\Pi_{\psi_A}^{\rm in}\otimes\Pi_{\phi_B}^{\rm in})\rho_{0}(\Pi_{\psi_A}^{\rm in}\otimes\Pi_{\phi_B}^{\rm in})]
\end{equation}
with $\rho_0$ the initial density operator.

\subsection{Quantum entropy distribution}
\label{appxB}

The fluctuations of the stochastic entropy production of a quantum system obey quantum fluctuation theorems\,\cite{Sagawa2014,Gherardini_entropy,ManzanoPRX,Irreversibility_chapter}. They are determined by evaluating the forward and backward processes associated to the dynamical evolution of the system. In particular, one can find that the stochastic quantum entropy production $\Delta\sigma_{m,n}$ equals to
\begin{equation}
\Delta\sigma(a^{\textrm{fin}}_{m},a^{\textrm{in}}_{k}) \equiv \ln\frac{p_{\rm F}(a^{\textrm{fin}}_{m},a^{\textrm{in}}_{k})}{p_{\rm B}(a^{\textrm{in}}_{k},a^{\textrm{ref}}_{m})} \ ,
\end{equation}
where $p_{\rm F}(a^{\textrm{fin}}_{m},a^{\textrm{in}}_{k})$ and $p_{\rm B}(a^{\textrm{in}}_{k}, a^{\textrm{ref}}_{m})$ are the joint probabilities to simultaneously measure the outcomes $\{a\}$ in a single realization of the forward $(\rm F)$ and backward $(\rm B)$ process, respectively. Instead, $\{a\}$ denotes the set of measurement outcomes from a generic TPM scheme in which the measurement observables $\mathcal{O}$ are not necessarily the system Hamiltonian at $t_0$ and $t_{\rm fin}$. However, in our case the measurement outcomes $\{a\}$ are chosen equal to the energies $\{E\}$ of the system. Then, as proved in Ref.\,\cite{Gherardini_entropy}, the outcome $a^{\textrm{ref}}_{m}$ refers to the state after the $1^{\rm st}$ measurement of the backward process, which is called reference state. If the evolution of the system is unital (in our case, the dynamics is simply unitary), then the stochastic quantum entropy production $\Delta\sigma_{m,n}$ becomes
\begin{equation}
\Delta\sigma(E^{\textrm{fin}}_{m},E^{\textrm{in}}_{k}) = \ln\frac{p(E^{\textrm{in}}_{k})}{p(E^{\textrm{ref}}_{m})} \ ,
\end{equation}
with $p(E^{\textrm{ref}}_{m})$ denoting the probability to get the measurement outcome $E_{m}^{\rm ref}$. Although the quantum fluctuation theorem can be derived without imposing a specific operator for the reference state\,\cite{Sagawa2014}, it is worth choosing the latter equal to the final density operator after the $2^{\rm nd}$ measurement of the forward process. This choice appears to be the most natural among the possible ones to design a suitable measuring scheme of general thermodynamic quantities, consistently with the quantum fluctuation theorem and the asymmetry of the second law of thermodynamics. This means that for our purposes the stochastic quantum entropy production is
\begin{equation}
\Delta\sigma(E^{\textrm{fin}}_{m},E^{\textrm{in}}_{k}) = \ln p(E^{\textrm{in}}_{k}) - \ln p(E^{\textrm{fin}}_{m}) \ ,
\end{equation}
where $p(E^{\textrm{fin}}_{m})$ denotes the probability to measure the $m^{\rm th}$ energy outcome at the final time instant $t_{\rm fin}$.

Experimentally, it is not possible in general to derive the stochastic realizations of the quantum entropy production of a multipartite quantum system by just performing local measurements. Specifically, it becomes feasible if the dynamical map of the composite system acts separately on each partition of the system and $\rho_0$ is a product state. 

\begin{figure}[t!]
    \centering
    \includegraphics[width = 0.975\columnwidth]{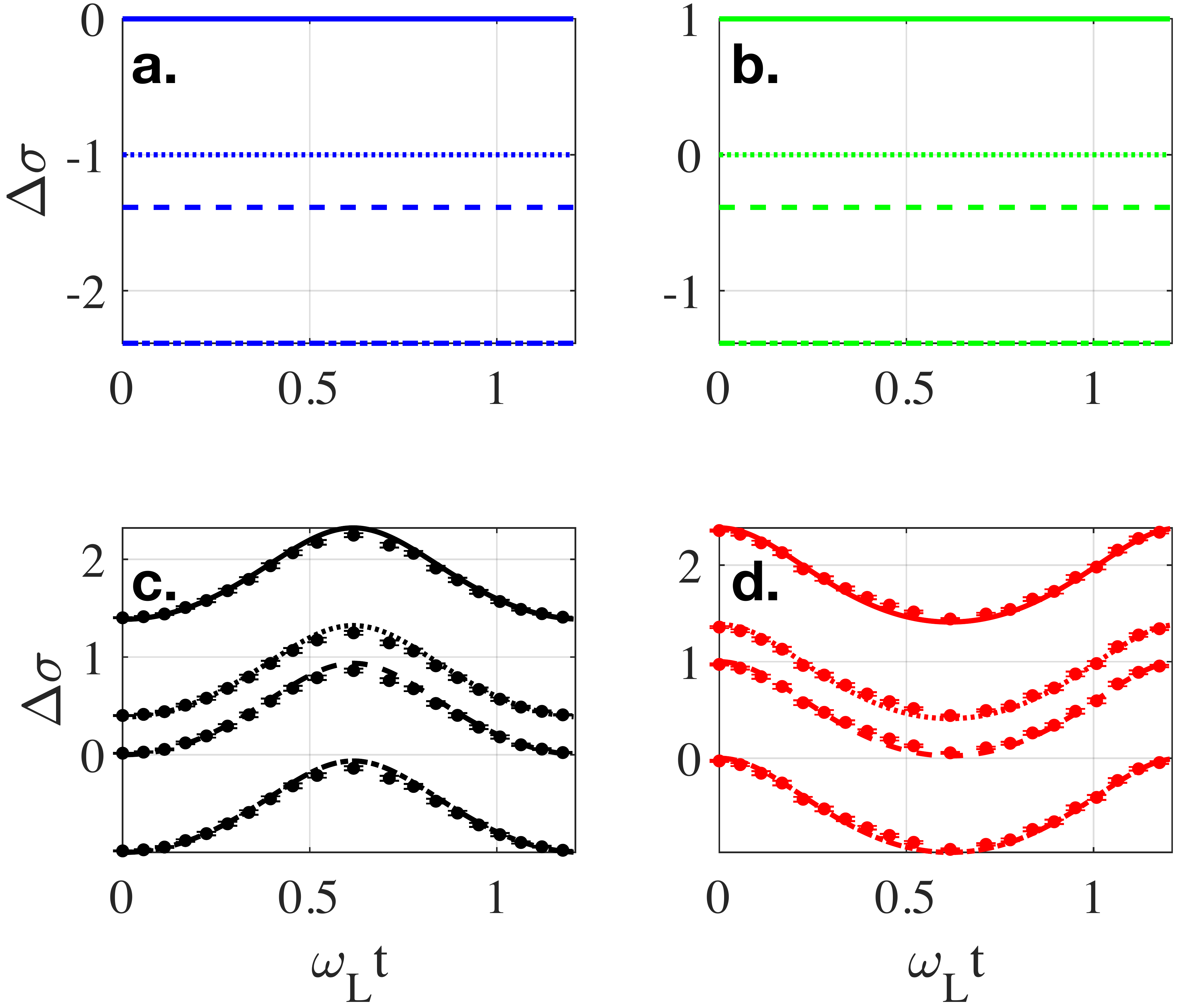}
    \caption{{\bf Realizations of $\Delta\sigma$}. Theoretical (lines) and experimental (dots)} realizations of $\Delta\sigma$ as a function of time. As in the other figures, $\omega_{\rm int}/\omega_{L} = 5$ in natural units and $\rho_0$ is provided by Eq.\,(\ref{eq_rho_0}). In each inset the values assigned to the elements of the set $(\psi_{A}^{\rm in},\phi_{B}^{\rm in},\psi_{A}^{\rm fin},\phi_{B}^{\rm fin})$ are respectively given as: a. Blue solid line: $(0_A,0_A,0_B,0_B)$, blue dotted line: $(0_A,1_B,0_B,0_B)$, blue dashed line: $(1_A,0_B,0_B,0_B)$, blue dash-dotted line: $(1_A,1_B,0_B,0_B)$; b. Green solid line: $(0_A,0_A,0_B,1_B)$, green dotted line: $(0_A,1_B,0_B,1_B)$, green dashed line: $(1_A,0_B,0_B,1_B)$, green dash-dotted line: $(1_A,1_B,0_B,1_B)$; c. Black solid line: $(0_A,0_A,1_B,0_B)$, black dotted line: $(0_A,1_B,1_B,0_B)$, black dashed line: $(1_A,0_B,1_B,0_B)$, black dash-dotted line: $(1_A,1_B,1_B,0_B)$; d. Red solid line: $(0_A,0_A,1_B,1_B)$, red dotted line: $(0_A,1_B,1_B,1_B)$, red dashed line: $(1_A,0_B,1_B,1_B)$, red dash-dotted line: $(1_A,1_B,1_B,1_B)$.
    \label{fig:stochastic_entropy}
\end{figure}

In such a case, indeed, one can write
\begin{equation}
\rho_{\rm in} = \sum_{n}p(E^{\textrm{in}}_{n})\Pi_{n}^{\rm in}
\end{equation}
and
\begin{equation}
\begin{aligned}
p(E^{\textrm{fin}}_{m})&= \sum_{n}p(E^{\textrm{in}}_{n}){\rm Tr}\left[\Lambda_{t_{\rm fin}}[\Pi_{n}^{\rm in}]\Pi_{m}^{\rm fin}\right]\\
&= \sum_{n}p(E^{\textrm{fin}}_{m}|E^{\textrm{in}}_{n})p(E^{\textrm{in}}_{n}),
\end{aligned}
\end{equation}
where $\Pi_{n}^{\rm in} \equiv \Pi_{A}^{\rm in} \otimes \Pi_{B}^{\rm in}$. Thus, by experimentally measuring the conditional probabilities $p(E^{\textrm{fin}}_{m}|E^{\textrm{in}}_{n})$ and the set $\{p(E^{\textrm{in}}_{n})\}$, one can also determine the set $\{p(E^{\textrm{fin}}_{m})\}$ of final probabilities, as well as the $16$ realizations of the stochastic quantum entropy production -- see Fig.\,\ref{fig:stochastic_entropy}. In Fig.\,\ref{fig:stochastic_entropy}, one can observe that only the black and red lines are not constant in time: they all correspond to the situation of finding qubit $A$ at $t_{\rm fin}$ in the eigenstate $|1\rangle_A$ of the local Hamiltonian $\mathcal{H}_{L_{A}}$.

Now, let us introduce the probability distribution $\textrm{Prob}(\Delta\sigma)$. Depending on the values assumed by the measurement outcomes $\{E^{\textrm{in}}\}$ and $\{E^{\textrm{fin}}\}$, $\Delta\sigma$ is a fluctuating variable. Thus, each time we repeat the TPM scheme, we have a different realization for $\Delta\sigma$ within a set of discrete values. The probability distribution $\textrm{Prob}(\Delta\sigma)$ is fully determined by the knowledge of the measurement outcomes and the respective probabilities. As proved in\,\cite{Gherardini_entropy,ManzanoPRX}, it is equal to
\begin{equation*}
\textrm{Prob}(\Delta\sigma) = \sum_{k,m}\delta\left(\Delta\sigma - \Delta\sigma(E^{\textrm{in}}_{k},E^{\textrm{fin}}_{m})\right)p(E_{k}^{\rm in},E_{m}^{\rm fin}).
\end{equation*}
Once again, it is worth noting that the specific values of $E_{k}^{\rm in}$ and $E_{m}^{\rm fin}$ change whether we apply global or local energy measurements on the bipartite quantum system.

\subsection{Comparison between energy change and entropy distributions}

\begin{figure}[t!]
    \centering
    \includegraphics[width=0.98\columnwidth]{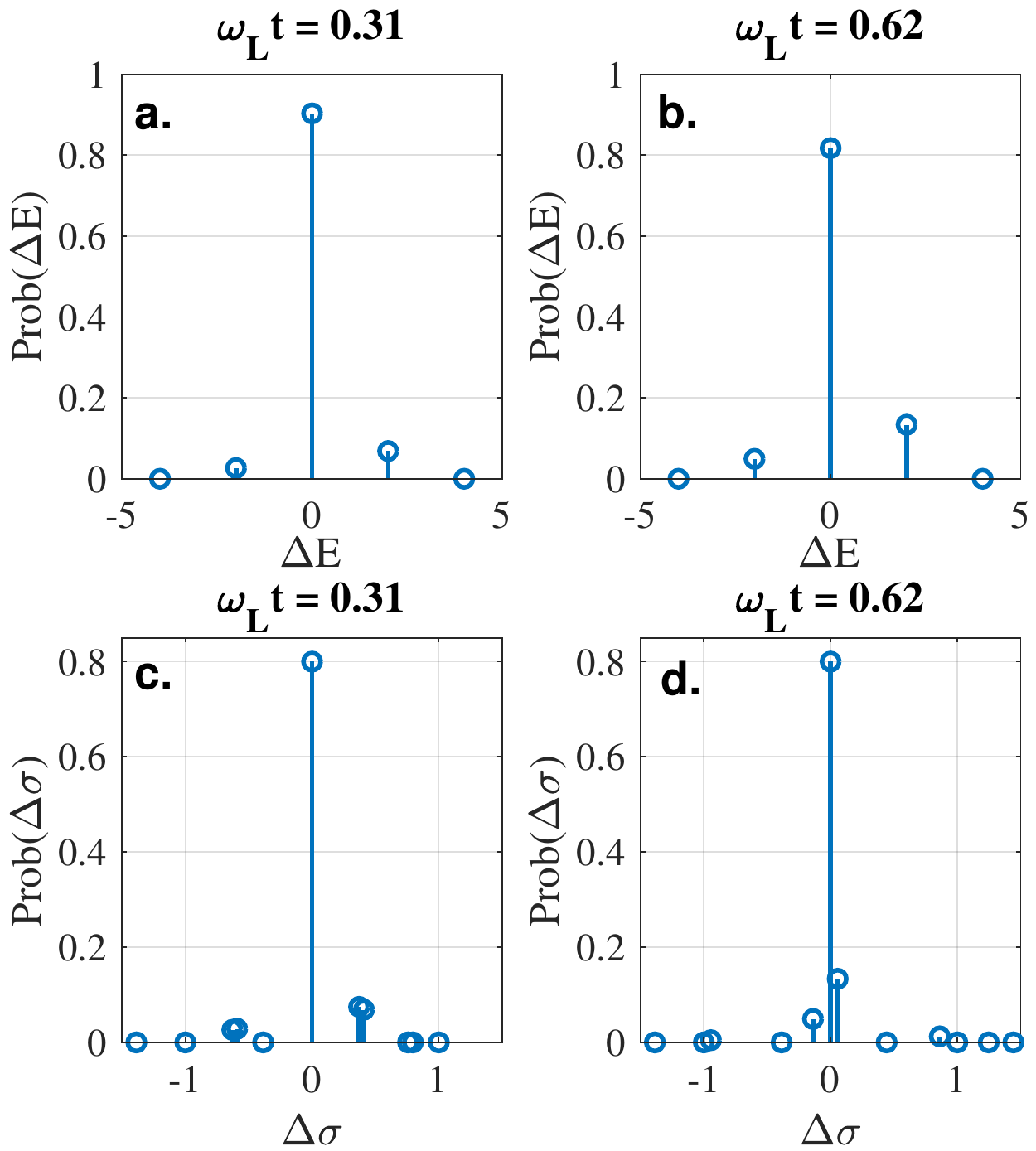}
    \caption{{\bf Energy and entropy distributions}. Comparison between the experimental distributions of $\Delta E$ [panels a.\,and b.] and $\Delta\sigma$ [panels c.\,and d.], evaluated at $\omega_{L}t= 0.31$ and $0.62$. It is worth noting that in such (rescaled) time instants the first statistical moments (i.e., the average) of the energy change and the entropy, respectively, take their maximum values.}
    \label{fig:distr_joint}
\end{figure}

The trends followed by the statistical moments of the energy changes and entropy production is markedly different. While the moments of $\Delta E$ (up to those reported in this work) are all positive within the time window that we have addressed, $\Delta\sigma^{3}$ and $\Delta\sigma^{5}$ can take negative values. These features have implications in the shape taken by the respective probability distributions, which are reported in Fig.\,\ref{fig:distr_joint} for two choices of the rescaled time: the distribution of energy changes showcases a larger skewness with a short and fat right tail. The indefinite signs taken by the moments of the entropy production, on the other hand, keep the corresponding distribution very symmetric around $\Delta\sigma=0$.

\subsection{Error analysis}

\begin{figure}[t!]
    \centering
    \includegraphics[width = \columnwidth]{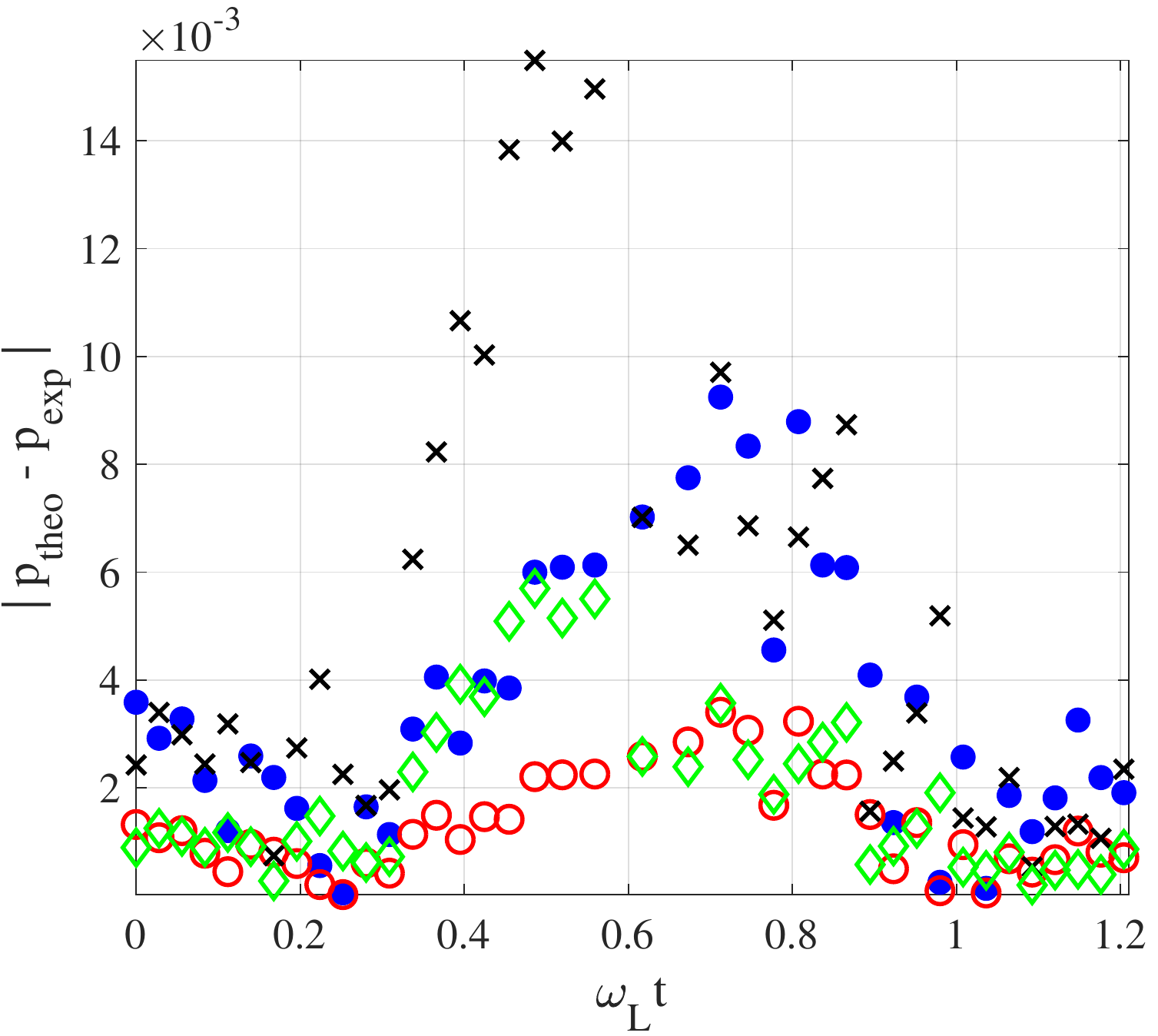}
    \caption{{\bf Error analysis}. Absolute error $|p_{\rm theo}-p_{\rm exp}|$ accounting for the difference between theoretical and experimental joint probabilities $p(E_{1_{A}\phi_{B}}^{\rm in},E_{1_{A}\phi_{B}}^{\rm fin})$ as a function of the rescaled time $\omega_Lt$. The blue dots report the values for $p(E^{\rm in}_{1_{A}0_{B}},E^{\rm fin}_{1_{A}0_{B}})$, the red circles those for $p(E^{\rm in}_{1_{A}1_{B}},E^{\rm fin}_{1_{A}0_{B}})$,  the black crosses and the green diamond label the values for $p(E^{\rm in}_{1_{A}0_{B}},E^{\rm fin}_{1_{A}1_{B}})$ and $p(E^{\rm in}_{1_{A}1_{B}},E^{\rm fin}_{1_{A}1_{B}})$, respectively. As in Fig.\,\ref{fig:joint_prob}, we have taken $\omega_{\rm int}/\omega_L = 5$. The initial state $\rho_0$ is given by Eq.\,(\ref{eq_rho_0}) and $|V\rangle_k=(1,0)$, $|H\rangle_k=(0,1)$ for both qubits $k=A$ and $B$.}
    \label{fig:rel_err_joint}
\end{figure}

In Fig.\,\ref{fig:rel_err_joint} we report the absolute difference between the theoretical and experimental joint probabilities for various initial-final configurations of the system, against the rescaled time $\omega_L t$. The trend followed by such discrepancies is consistent across the various initial-final configurations that we have considered, with larger values showcases close to $\omega_L t \simeq 0.62$.

\subsection{Data availability}
Data are available to any reader upon reasonable request.

\section*{Acknowledgments}
\acknowledgments 
The authors thank A.\,Belenchia for a critical reading of the manuscript and useful comments. 
SG, LB and FC were financially supported by the Fondazione CR Firenze through the project Q-BIOSCAN and QUANTUM-AI, PATHOS EU H2020 FET-OPEN grant No.\,828946, and UNIFI grant Q-CODYCES. SG also acknowledges the MISTI Global Seed Funds MIT-FVG grant program. MP gratefully acknowledge support by the H2020 Collaborative Project TEQ (Grant Agreement 766900), the SFI-DfE Investigator Programme through project QuNaNet (grant 15/IA/2864), the Leverhulme Trust through the Research Project Grant UltraQuTe (grant nr.\,RGP-2018-266) and the Royal Society through the Wolfson Fellowship scheme (RSWF\textbackslash R3\textbackslash183013) and the International Exchange scheme (grant number IEC\textbackslash R2\textbackslash192220).

\section*{Author contributions}
SG, MB and MP conceived the original idea. SG and MB performed the initial calculations which were then developed with the help of LB, VC, MP and FC. VC led the experimental endeavours, with assistance by IG and MS. All authors discussed the results and their interpretation. SG, VC, MP and MB wrote the manuscript with input from all the other authors. MP, MB and FC supervised the project.

\section*{Competing interests}
The authors declare no competing interests.

\end{document}